\documentclass{article}
\renewcommand{\epsilon}{\varepsilon}
\begin{document}

\title{On homothetic cosmological dynamics}
\author{Rinat A. Daishev\footnote{Electronic address:
Rinat.Daishev@ksu.ru} \\
Department of General Relativity and Gravitation  \\
Kazan State University, 420008 Kazan, Russia\\
and\\
Winfried Zimdahl\footnote{Electronic address:
zimdahl@thp.uni-koeln.de}\\
Fachbereich Physik, Universit\"at Konstanz\\ PF M678, D-78457
Konstanz, Germany\\and\\
Institut f\"ur Theoretische Physik, Universit\"at zu K\"oln\\
Z\"ulpicher Str. 77, D-50937 K\"oln, Germany }

\date{\today}

\maketitle

\begin{abstract}
We consider the homogeneous and isotropic cosmological fluid
dynamics which is compatible with a homothetic, timelike motion,
equivalent to an equation of state $\rho + 3P = 0$. By splitting
the total pressure $P$ into the sum of an equilibrium part $p$ and
a non-equilibrium part $\Pi$, we find that on thermodynamical
grounds this split is necessarily given by $p=\rho$ and $\Pi =
-\frac{4}{3}\rho$, corresponding to a dissipative stiff
(Zel'dovich) fluid.
\end{abstract}

\section{Introduction}

Homothetic motions and self-similar spacetimes have been broadly
discussed in the literature, mainly from a mathematical but also
from a physical point of view (for a recent review see
\cite{CaCo}). Applications in astrophysics and cosmology in many
cases rely on perfect fluid matter models, but dissipative fluids
have been investigated as well \cite{BuCo}. Here, we are
interested in a specific aspect of the cosmological dynamics,
namely in the imperfect fluid equations of state  which are
compatible with the existence of a timelike homothetic vector.

Timelike symmetries are not generally expected to exist in the
expanding universe. A timelike  Killing vector  characterizes a
stationary spacetime. But it is known \cite{TauWei,Coley} that
under certain circumstances a {\it conformal}, timelike symmetry
is possible in a Friedmann-Lema\^{\i}tre-Robertson-Walker (FLRW)
universe. The existence of a conformal timelike Killing vector is
closely related to ``global'' equilibrium properties of perfect
fluids. For massless particles the condition for global
equilibrium requires the quantity $u ^{a}/T$, where $u^a$ is the
fluid 4-velocity and $T$ is the fluid temperature, to be a
conformal Killing vector (CKV). This CKV is compatible with the
cosmological expansion for an ultra-relativistic equation of
state. Any deviation from the latter will destroy the conformal
symmetry. In other words, the conformal symmetry singles out a
perfect fluid with the equation of state for radiation.
Apparently, this also implies that a conformal symmetry is
incompatible with a non-vanishing entropy production. Only the
conformal symmetry of an ``optical'' metric, in which an effective
refraction index of the cosmic substratum characterizes specific
internal interactions that macroscopically correspond to a
negative pressure contribution, may be compatible with the
production of entropy \cite{ZBPRD,ZBENT,moscow}. In the present
paper we want to point out a  different feature of the connection
between symmetries and entropy production which, however, also
relies on a dissipative scalar pressure.

The cosmological principle restricts the  energy momentum tensor
$T^{ik}$ of the cosmic substratum to be of the structure
\begin{equation}
T ^{ik}= \rho  u ^{i}u ^{k} + P h ^{ik}\ . \label{1}
\end{equation}
Here, $\rho $ is the energy density measured by an observer
comoving with the fluid four-velocity $u ^{i}$ which is normalized
by $u ^{a}u _{a} = -1$. The quantity $h _{ik} \equiv  g _{ik} + u
_{i}u _{k}$ is the spatial projection tensor with $h _{ik}u ^{k} =
0$. The total pressure $P$ is the sum
\begin{equation}
P = p + \Pi \label{2}
\end{equation}
of an equilibrium part $p>0$ and a non-equilibrium part $\Pi \leq
0$ which is connected with entropy production. A perfect fluid is
characterized by $\Pi =0$. A scalar, viscous pressure is the only
entropy producing phenomenon which is compatible with  spatial
homogeneity and isotropy. The idea of our paper is to impose a
homothetic symmetry requirement on the dynamics of this medium and
to ask for the admissible equations of state for $p$ and $\Pi$. To
this purpose we shall combine the homothetic, timelike dynamics in
a Friedmann-Lema\^{\i}tre-Robertson-Walker (FLRW) universe with
thermodynamic considerations within the latter. It will turn out
that for $P = \alpha \rho$ and $p = w \rho$, where $\alpha$ and
$w$ are constants, we have necessarily $\alpha = -1/3$ (cf.
\cite{Eardley,CaCo}) and the split $P=p+\Pi$ is uniquely given by
$w=1$, corresponding to a dissipative fluid with $p = \rho$ and
$\Pi = -(4/3)\rho$. An equation of state $P = -1/3\rho$ has been
associated with cosmic string matter \cite{Vilenkin}, while
$p=\rho$ characterizes a stiff (Zel'dovich) fluid. This means,
under the condition of homothetic symmetry a substance with $P =
-1/3\rho$ is dynamically equivalent to dissipative stiff matter.
This feature of having two interpretations for the same overall
equation of state is a property of the homothetic symmetry only
and does not hold for a general conformal symmetry which singles
out an equation of state $P= p= \rho /3$, i.e., $\Pi =0$.

In section \ref{Thermodynamic relations} we briefly recall
relevant thermodynamic relations. Section \ref{Conformal symmetry}
is devoted to the conformal symmetry in the expanding universe
while the homothetic motion is characterized in section
\ref{Homothetic motion}. Section \ref{Conclusions} summarizes our
results.

\section{Thermodynamic relations}
\label{Thermodynamic relations}

In this section we recall basic thermodynamic relations which will
be relevant for the symmetry considerations below. Energy-momentum
conservation $T^{ik}_{;k}=0$ provides us with
\begin{equation}
\dot{\rho} +\Theta\left(\rho + P \right) = 0\ , \label{3}
\end{equation}
where the quantity $\Theta \equiv  u ^{i}_{;i}$ is the fluid
expansion and $\dot{\rho} \equiv  \rho_{,i}u ^{i}$. The particle
number balance is
\begin{equation}
N ^{i}_{;i} = \dot{n} + \Theta n = n \Gamma \ , \label{4}
\end{equation}
where $N^i = n u^i$ is the particle number flow vector and $n$ is
the particle number density. For $\Gamma > 0$ we have particle
creation. $\Gamma = 0$ corresponds to particle number
conservation. The possible origin of a non-vanishing particle
production rate $\Gamma$, e.g., a phenomenological description of
quantum particle production out of the gravitational field
\cite{Zel,Mur,Hu,Turok,Barr,Prig} is not relevant here. In the
present context the production rate $\Gamma$ is {\it not} given by
by microphysical considerations but it follows from the
(homothetic) symmetry condition (see below). The imposed symmetry
may require a certain production rate to be realized at all.
Alternatively, as we shall see, the requirement $\Gamma =0$ under
the condition of a homothetic motion leads to an increase in the
entropy per particle. Dynamically, both effects are equivalent.
From the Gibbs equation
\begin{equation}
T \mbox{d}s = \mbox{d} \frac{\rho }{n} + p \mbox{d}\frac{1}{n}\ ,
\label{5}
\end{equation}
where $s$ is the entropy per particle, it follows that
\begin{equation}
n T \dot{s} = \dot{\rho } - \left(\rho + p \right)
\frac{\dot{n}}{n}\ . \label{6}
\end{equation}
Using here the balances (\ref{3}) with (\ref{2}) and (\ref{4})  yields
\begin{equation}
n T \dot{s} = - \Theta \Pi - \left(\rho + p \right)\Gamma \ .
\label{7}
\end{equation}
From the Gibbs-Duhem relation
\begin{equation}
\mbox{d} p = \left(\rho + p \right)\frac{\mbox{d} T}{T} + n T
\mbox{d} \left(\frac{\mu }{T} \right)\ , \label{8}
\end{equation}
where $\mu $ is the chemical potential, we obtain
\begin{equation}
\left(\frac{\mu }{T}\right)^{^{\displaystyle \cdot}} =
\frac{\dot{p}}{nT} - \frac{\rho + p}{nT}\frac{\dot{T}}{T}\ .
\label{9}
\end{equation}
The above set of equations has to be supplemented by
(explicitly not yet known)
equations of state which are assumed to have the general form
\begin{equation}
p = p \left(n,T \right)\ ,\ \ \ \ \ \ \ \ \ \rho = \rho \left(n,T
\right)\ , \label{10}
\end{equation}
i.e., particle number density and temperature are taken as the
independent thermodynamical variables. Differentiating the latter
relation and using the balances (\ref{3}) and (\ref{4}) and 
relation (\ref{7}) provides
us with an evolution law for the temperature:
\begin{equation}
\frac{\dot{T}}{T} = - \left(\Theta - \Gamma  \right)
\frac{\partial p}{\partial \rho } + \frac{n \dot{s}}{\partial \rho
/ \partial T}\ , \label{11}
\end{equation}
where the abbreviations
\[
\frac{\partial{p}}{\partial{\rho }} \equiv
\frac{\left(\partial p/ \partial T \right)_{n}}
{\left(\partial \rho / \partial T \right)_{n}} \ ,
\ \ \ \ \ \ \ \
\frac{\partial{\rho }}{\partial{T}} \equiv
\left(\frac{\partial \rho }{\partial T} \right)_{n}\ ,
\]
have been used, as well as the general relation
\begin{equation}
\frac{\partial{\rho }}{\partial{n}} = \frac{\rho + p}{n} -
\frac{T}{n}\frac{\partial{p}}{\partial{T}}\ , \label{12}
\end{equation}
which follows from the fact that the entropy is a state function.
Together with the particle number balance (\ref{4})  the entropy
flow vector $S ^{a} = ns u ^{a}$ gives rise to the following
expression for the entropy production density:
\begin{equation}
S ^{a} = ns u ^{a} \quad\Rightarrow\quad S ^{a}_{;a}=ns \Gamma  +
n \dot{s}\ .
\label{13}
\end{equation}
Entropy may be produced either by an increase in the number of
particles or by an increase in the entropy per particle. Both
cases will play a role in our analysis.

\section{Conformal symmetry}
\label{Conformal symmetry}

The condition for the quantity $\xi^i \equiv u^i /T$ to be a CKV
is
\begin{equation}
\pounds _{\xi } g _{ik} =  \phi g _{ik}   \ . \label{14}
\end{equation}
This condition implies
\begin{equation}
\phi =  \frac{{\rm 2}}{{\rm 3}} \frac{{\rm \Theta } }{T}\
\label{15}
\end{equation}
and
\begin{equation}
\frac{\dot{T}}{T} = - \frac{\Theta}{3}\ . \label{16}
\end{equation}
In the homogeneous and isotropic case under consideration here,
the Raychaudhuri equation for the expansion scalar reduces to
\begin{equation}
\dot{\Theta} + \frac{1}{3}\Theta^2 + \frac{\kappa}{2}\left(\rho +
3P\right) = 0 \ , \label{17}
\end{equation}
with $\kappa = 8\pi G$, where $G$ is Newton's gravitational constant. 
The derivative of $\phi$ in (\ref{15}) then becomes
\begin{equation}
\dot{\phi} = - \frac{\kappa}{3}\frac{\rho + 3P}{T}  \ . \label{18}
\end{equation}
Combining the general temperature law (\ref{11}) with relation
(\ref{16}) yields
\begin{equation}
\Gamma + \frac{n \dot{s}}{\partial p / \partial T}= \left(1 -
\frac{1}{3}\frac{\partial \rho}{\partial p}\right)\Theta \ ,
\label{19}
\end{equation}
with $\Theta$ being related to the conformal factor by  Eq.
(\ref{15}).
\ \\
Assuming now an overall equation of state $P=\alpha \rho$, the
Lie-derivatives of the relevant quantities become
\begin{equation}
\pounds _{\xi} \rho =  - \frac{3}{2}\left(1+\alpha\right)\phi\rho
\ , \qquad \pounds _{\xi} P =  -
\frac{3}{2}\left(1+\alpha\right)\phi P
 \ ,
 \label{20}
\end{equation}
\begin{equation}
\pounds _{\xi } u _{i} =   \frac{1}{2}\phi u _{i}   \ , \qquad
\pounds _{\xi } u ^{i} =   - \frac{1}{2}\phi u^{i}   \ ,
\label{21}
\end{equation}
and
\begin{equation}
\pounds _{\xi} T _{ab}=  - \frac{1}{2}\left(1+3\alpha\right)\phi
T_{ab} \ .
 \label{22}
\end{equation}
Taking into account that (for $\mu =0$)
\begin{equation}
s =  \frac{\rho + p}{nT}\ , \label{23}
\end{equation}
and assuming additionally $p=w\rho$ with a constant $w$, we have
also
\begin{equation}
\pounds _{\xi} s =  - \frac{3}{2}\frac{\alpha - w}{1+w}  \phi s
 \ .
 \label{24}
\end{equation}
For the standard case $\Gamma =\dot{s} = 0$ the relation
(\ref{19}) is satisfied for $p=\rho /3$. This corresponds to the
well-known fact that for a perfect fluid the conformal symmetry
condition (\ref{14}) singles out ultra-relativistic matter.
However, one may also ask the inverse question, namely: Does the
assumption of a total equation of state $P=\rho/3$ necessarily
imply $\Pi=0$? We shall investigate this for the cases $\dot{s}
=0$ and $\Gamma =0$ separately.  For the case $\dot{s} =0$ we
obtain from Eq. (\ref{7}) that (cf. \cite{Prig,Calv,ZP})
\begin{equation}
\Gamma = - \frac{\Pi}{\rho + p}\Theta \ \label{25}
\end{equation}
is valid which, together with (\ref{19}), leads to
\begin{equation}
- \Pi = \left(1 - \frac{1}{3}\frac{\partial \rho}{\partial
p}\right)\left(\rho + p\right) \ , \label{26}
\end{equation}
or
\begin{equation}
\rho + P = \frac{1}{3}\frac{\partial \rho}{\partial p}\left(\rho +
p\right) \ . \label{27}
\end{equation}
Introducing here $P=\rho /3$ and $p=w\rho$, relation (\ref{27})
can only be satisfied for $w=1/3$, which indeed is equivalent to
$\Pi=0$. For the second special case $\Gamma =0$ we have from
(\ref{7})
\begin{equation}
n\dot{s} = - \frac{\Theta \Pi}{T} \ , \label{28}
\end{equation}
which combined with (\ref{19}) leads to
\begin{equation}
- \Pi = T \frac{\partial }{\partial T} \left[p -
\frac{1}{3}\rho\right] \ . \label{29}
\end{equation}
With $P=\rho /3$ and $p=w\rho$, as well as with Eq. (\ref{12}) we
find that $\rho \propto nT$. Applying additionally the Gibbs-Duhem
equation  (\ref{8}) and the second equation of state in
(\ref{10}), we end up again with $w=\alpha = 1/3$ and $\Pi=0$.
Consequently, in none of these cases the conformal symmetry is
compatible with a dissipative fluid configuration.

\section{Homothetic motion}
\label{Homothetic motion}

Now we assume the conformal factor to be constant, i.e., a
homothetic motion. According to (\ref{18}) this implies
\begin{equation}
\phi = {\rm const} \quad \Rightarrow \quad \rho + 3P=0 \ .
\label{30}
\end{equation}
An equation of state $P = - \rho /3$ characterizes cosmic string
matter \cite{Vilenkin}. The Raychaudhuri equation (\ref{17}) in
such a case reduces to
\begin{equation}
\dot{\Theta} + \frac{1}{3}\Theta^2 = 0 \ . \label{31}
\end{equation}
Introducing here the scale factor $a$ of the Robertson-Walker metric
by
\begin{equation}
\Theta \equiv 3H \equiv 3 \frac{\dot{a}}{a}\ , \label{32}
\end{equation}
where $H$ is the Hubble-parameter, results in
\begin{equation}
\Theta \propto \frac{1}{a}\quad \Rightarrow \quad a \propto t \ .
\label{33}
\end{equation}
Under the homothetic condition the energy balance (\ref{3}) yields
\begin{equation}
\rho \propto \frac{1}{a^2} \ . \label{34}
\end{equation}
Since from (\ref{16}) with (\ref{32}) one has $T \propto a^{-1}$,
this implies $\rho \propto T^2 $. The relations (\ref{16}),
(\ref{32}) and (\ref{33}) are consistent  with $\phi \propto
\Theta/ T = {\rm const}$. Furthermore, (\ref{33}) and (\ref{34})
are consistent with Friedmann's equation
\begin{equation}
8\pi G\rho = \frac{\Theta^2}{3}  + 3 \frac{k}{a^2}\ , \label{35}
\end{equation}
where $k=0,\pm 1$. Again we consider the cases $\dot{s}=0$ and
$\Gamma =0$ separately. From Eq.(\ref{27}) (valid for $\dot{s}=0$)
we obtain
\begin{equation}
\frac{\partial \rho}{\partial p}\left(1 + \frac{p}{\rho}\right) =
2 \ .
\label{36}
\end{equation}
In this case $p = w \rho$ leads to $w=1$, i.e., the equilibrium
equation of state is that for stiff matter. Consequently, we have
\begin{equation}
\Pi = -\frac{4}{3}\rho \ . \label{37}
\end{equation}
This means, different from the conformal case, {\it the homothetic
symmetry is compatible with a dissipative fluid configuration}.
The set (\ref{20}) - (\ref{22}) of Lie-derivatives becomes
\begin{equation}
\pounds _{\xi} \rho =  - \phi\rho \ , \qquad \pounds _{\xi} P =  -
\phi P\ , \qquad \pounds _{\xi} T _{ab}=  0 \ ,\qquad \pounds
_{\xi} s =  \phi s
 \ .
 \label{38}
\end{equation}
From (\ref{29}) with $-\Pi =p + \frac{\rho}{3}$ and $\rho \propto
T^2$ we find $p=\rho$ also in the case $\Gamma =0$, $\dot{s} \neq
0$. This means, Eq. (\ref{37}) is valid for both the cases
considered here. For $\dot{s} = 0$ one has
\begin{equation}
\Gamma = 2H= \phi T \ , \label{39}
\end{equation}
i.e., the production rate is required to be twice the Hubble rate
$H$. For the case $\Gamma = 0$ it follows
\begin{equation}
\frac{\dot{s}}{s}= 2H
\label{40}
\end{equation}
for the fractional change of the entropy per particle. To complete
the thermodynamical description of our system we realize that $\Pi
= - 4/3\rho$ in (\ref{29}) amounts to
\begin{equation}
\frac{4}{3}\rho = \frac{2}{3} T \frac{\partial \rho}{\partial T}
\quad \Rightarrow \quad  T \frac{\partial \rho}{\partial T} =
2\rho \ , \label{41}
\end{equation}
which is consistent with $\rho \propto T^2$. Using the last
equation of (\ref{41}) in the expression (\ref{12}) for $\partial
\rho/\partial n$ we find
\begin{equation}
\frac{\partial \rho}{\partial n} = 0 \ . \label{42}
\end{equation}
All this is consistent with the Gibbs-Duhem equation (\ref{8}) for
$\mu =0$ since
\begin{equation}
\mbox{d} p = \left(\rho + p \right) \frac{\mbox{d} T}{T} \quad
\Rightarrow \quad \mbox{d} \rho = 2\rho \frac{\mbox{d} T}{T} \ .
\label{43}
\end{equation}
If we formally define a coefficient of bulk viscosity according
to $\Pi = - \zeta \Theta $, we obtain
\begin{equation}
\zeta = \frac{8}{9}\frac{\rho}{\phi T}
= \frac{m_P^2}{6\pi}H\left[1 + \frac{k}{a^2 H^2}\right]\ .
\label{44}
\end{equation}
Here we have applied equations (\ref{37}), (\ref{15}) and 
(\ref{35}) and introduced the square of the Planck mass 
$m_P^2 =G^{-1}$. Because of the dependence (\ref{33}) one has 
$a^2 H^2 = {\rm const}$ in the last expression of (\ref{44}). 

\section{Conclusion}
\label{Conclusions}

Given that the total pressure $P$ of the homogeneous and isotropic
cosmic medium  splits into an equilibrium part $p$ and a
non-equilibrium contribution $\Pi$,  and at the same time $u^i/T$
is a homothetic vector, the equations of state are
necessarily $p=\rho$ and $\Pi =-(4/3)\rho$. This corresponds to a
dissipative Zel'dovich fluid with an effective bulk viscosity coefficient
$\zeta \propto \rho^{1/2}$.\\
\ \\
\ \\
{\bf Acknowledgments}\\
\ \\
This paper was supported by the Deutsche Forschungsgemeinschaft.

\end{document}